\documentclass[twocolumn,prb,aps,amssym,nofootinbib,floatfix,eqsecnum]{revtex4}
\usepackage{graphicx}
\newcommand{\be}{\begin{equation}}
\newcommand{\ee}{\end{equation}}
\newcommand{\bea}{\begin{eqnarray}}
\newcommand{\eea}{\end{eqnarray}}
\begin{document}
\title{Non-Fermi liquid behavior in Kondo Models\footnote{submitted to the 
special issue of J. Phys Soc. Japan:
 ``Kondo effect -- 40 Years after the Discovery''}}
\author{Ian Affleck}
\affiliation{Department of Physics \& Astronomy, University of British Columbia, 
Vancouver, B.C., Canada, V6T 1Z1}
\date{\today}
\begin{abstract}
Despite the fact that the low energy behavior of the basic Kondo model cannot 
be studied perturbatively it was eventually shown by Wilson, Anderson, Nozi\`eres 
and others to have a simple ``local Fermi liquid theory'' description. That 
is  electronic degrees of freedom become effectively non-interacting 
in the zero energy limit. 
However, 
generalized versions of the Kondo model involving more than 
one channel or impurity may exhibit low energy behavior of a less 
trivial sort which can, nonetheless, be solved exactly using 
either Bethe ansatz or conformal field theory 
and bosonization techniques. Now the low energy limit  
exhibits interacting many body behavior. For example, processes in which a single electron 
scatters off the impurity into a multi electron-hole state have 
a non-vanishing (and sometimes large) amplitude at zero energy. 
 This corresponds to a rare solveable example 
of non-Fermi liquid behavior. Essential features of these phenomena are reviewed here. 
\end{abstract}
\maketitle
\section{Introduction}
Kondo's seminal paper of forty years ago\cite{Kondo}
  showed that the low energy behavior of the Kondo 
model is fundamentally non-perturbative. 
I write the Hamiltonian as:
\begin{equation}
H=\sum_{\vec{k}\alpha}\psi^{\dagger\alpha}_{\vec{k}}
\psi_{\vec{k}\alpha}\epsilon(k)+J\vec{S}\cdot\sum_{\vec{k}\vec{k'}}
\psi^{\dagger}_{\vec k} \frac{\vec{\sigma}}{2}\psi_{\vec{k'}}
\label{hambas}
\end{equation}
 where $\psi_{\vec{k}\alpha}$'s are conduction electron
annihilation operators, (of momentum  $\vec{k}$, spin $\alpha$) and
$\vec{S}$ represents the spin of the  magnetic impurity with
\be [S^a, S^b]=i\epsilon^{abc}S^c.\ee
There is an implicit sum over electron spin indices, $\alpha$ in 
the interaction term in Eq. (\ref{hambas}).
The dimensionless coupling constant is 
\be \lambda \equiv J\nu,\ee
where $\nu$ is the density of states at the Fermi surface. 
As Kondo showed, perturbation theory in $\lambda$ is infrared divergent at low $T$.
  For instance, the temperature-dependent resisitivity for a dilute array of impurities is 
given by a formula of the form:
\begin{equation}
\rho(T)\sim[\lambda+\lambda^2 \ln \frac{D}{T}+...]^2
\end{equation} Here $D$ is the band-width.
No matter how small the coupling constant, $\lambda$, 
the higher order terms eventually overwhelm the lower 
order ones at low enough temperature. 
 This result stimulated an enormous amount of theoretical
work. As Nozi\`eres put it, ``Theorists  `diverged' on their own,
leaving the experiment realities way behind''.\cite{Nozieres1}
As was realized  later, the divergence of the resistivity formula has 
an elegant interpretation in terms of renormalization 
group (RG) concepts. The scale dependent effective coupling 
constant, $\lambda (T)$ diverges as $T\to 0$:
\be \lambda (T) \approx \lambda+\lambda^2 \ln \frac{D}{T}+\ldots \ee
The temperature at which the higher order terms overwhelm 
the lower order ones,
\be T_K\approx D\exp [-1/\lambda ],\label{T_Kdef}\ee
defines a fundamental energy scale. Perturbation theory 
can be applied for $T>>T_K$ but not for $T\leq T_K$. 
The low $T$ behavior is fundamentally non-perturbative.

Nonetheless, the physics was eventually shown by Wilson,\cite{Wilson}
 Anderson,\cite{Anderson} Nozi\`eres\cite{Nozieres2} 
and others to be  simple at {\it very} low energies, 
$E<<T_K$.  Only the intermediate energy range where 
$E$ is $O(T_K)$ defies a simple description. 
This simplicity at very low energies 
arises from the fact, that in a certain sense 
to be made precise below, $\lambda (T)\to \infty$ at $T\to 0$. 
This infinite $\lambda$ behavior is actually quite simple. 
(I begin with the case of an $S=1/2$ impurity.) 
One electron forms a singlet with the impurity.  The 
remaining low energy electronic degrees of freedom 
feel an infinite repulsion from the screened impurity 
which corresponds to a $\pi /2$ phase shift in the s-wave 
channel.  The induced electron-electron interactions 
among these low energy degrees of freedom 
become increasingly unimportant as the energy scale 
decreases, corresponding to irrelevant interactions 
in the renormalization group sense. They 
lead to a simple dependence of physical 
quantities on $T$ (or other energy scales) which 
can be Taylor expanded in powers of $T/T_K$. 
From a renormalization group viewpoint, the 
low energy fixed point $(\lambda \to \infty$) 
is simply non-interacting 
electrons,  like the high-energy 
fixed point ($\lambda=0$) except for the 
removal of the impurity spin and the presence 
of a modified boundary condition at the impurity location 
corresponding to the phase shift.   The  
Kondo model thus provides a rare example of 
a renormalization group flow between two different 
fixed points, both of which are trivial.

It is perhaps surprising that simple modifications 
of the basic Kondo model can completely change 
this trivial low energy behavior.\cite{Nozieres3} The simplest 
such modification is to include several channels 
of electrons, changing the Hamiltonian  of
Eq. (\ref{hambas}) to:
\begin{equation}
H=\sum_{\vec{k}\alpha ,i}\psi^{\dagger\alpha i}_{\vec{k}}
\psi_{\vec{k}\alpha i}\epsilon(k)+J\vec{S}\cdot\sum_{\vec{k}\vec{k'}}
\psi^{\dagger i}_{\vec k} \frac{\vec{\sigma}}{2}\psi_{\vec{k'} i}
\label{hamk}
\end{equation}
Here $i$ labels the ``channels'' and runs from $1$ to $k$ (not 
to be confused with the momentum label, $\vec k$).
This fundamental change in low energy behavior 
was first demonstrated by Nozi\`eres and Blandin\cite{Nozieres3} by 
simple, intuitive RG arguments. Some aspects of 
these models were later solved for exactly using the 
Bethe ansatz.\cite{Andrei,Wiegmann}  This is a powerful method which 
allows calculations of thermodynamic quantities 
at all temperatures, enabling a study of the 
cross over between high energy and low energy fixed points. 
Later, Ludwig and I developed techniques to also 
calculate dynamical correlation functions, the resistivity 
and, in fact, {\it any} low energy property of these models. 
This amounted to an exact solution for the low energy 
fixed point which no longer has a simple Fermi liquid 
form. Since the interactions in these models only occur 
near the origin the low energy {\it bulk} behavior is always 
that of free electrons.  However, there is no 
quarantee that the behavior near the origin is 
also free electron like. This happens to be 
the case for the simplest Kondo model but is not 
so in general. What does turn out to be true 
in general is that the low energy behavior 
 is given by a free electron model with some 
sort of conformally invariant boundary condition (BC)
at the origin (that is, at the impurity location).
 In the simplest case this BC 
is of a trivial type which just corresponds to 
the phase shift mentioned above. In other cases, 
this BC encodes certain non-trivial 
electron-electron interactions which occur 
only near the origin. Indeed the problem 
of finding and studying such BC's turns out 
to be surprisingly rich but fortunately 
was solved in an elegant and general way 
by Cardy.\cite{Cardy1}  Ludwig and 
I\cite{Affleck1,Affleck2,Affleck3,Affleck4,Ludwig1,Affleck5,Ludwig2} 
were able to 
adapt Cardy's methods to generalized Kondo models. [For 
a much more extensive review of this work 
than is given here, see Ref. (\onlinecite{Affleck6}).]
Among low energy properties which can be solved 
for by these techniques are the S-matrix 
for low energy scattering off the impurity. In 
the single-channel model, an  incoming electron at energy
near $E_F$ (the Fermi energy) only scatters into 
a single outgoing electron at low $T$.  All 
multi-electron-hole scattering processes 
have amplitudes which vanish as the energy 
and $T\to 0$. This is just like what 
happens in standard Landau Fermi liquid theory 
for electron-electron scattering. In 
some of the multi-channel  
models the amplitude for these multi-electron-hole 
processes is non-zero (and is sometimes large) in 
the zero energy limit.  Thus, we speak of 
non-Fermi liquid fixed points. 

Shortly after the general  solution of the 
non-Fermi liquid Kondo fixed points using 
boundary condition techniques, was obtained, 
an alternative, simpler approach was developed\cite{Emery}  
by Emery and Kivelson for the particular, 
and most experimentally relevant case, of 
$k=2$, $S=1/2$.  This is based on 
standard bosonization methods, and the study 
of a particular anisotropic model which can 
be mapped onto non-interacting fermions.

In this brief review, I will focus
exclusively on the multi-channel Kondo model 
as a route to non-Fermi liquid behavior.  However, 
another rather well-studied generalization of 
the basic Kondo model which exhibits non-Fermi liquid 
behavior is the 2 impurity Kondo model.\cite{Affleck8}  Other 
generalizations with non-Fermi liquid behavior 
are also known. At this point in time, the multi-channel 
(in particular, 2 channel) Kondo model appears 
to offer the best hope for experimental realization. 

In the next section I review the perturbative results 
on the multi-channel Kondo model and the Nozi\`eres-Blandin 
analysis\cite{Nozieres3}  of the strong coupling limit which first 
showed that the low energy fixed point was not of Fermi 
liquid type when the number of channels, $k>2S$, where 
$S$ is the size of the impurity spin. In Sec. III 
I discuss the mapping onto a one-dimensional model 
which can be approximated as a Dirac fermion on 
the half-line with the impurity at the origin. I
then discuss the $\pi /2$ phase shift as a 
simple boundary condition.  In Sec. IV I sketch 
some aspects of Cardy's general theory of conformally 
invariant BC's and how it was applied to the 
multi-channel Kondo model. In Sec. V I discuss 
the stability of the non-Fermi liquid fixed point 
against various types of symmetry breaking interactions 
which may be present in the Hamiltonian. I also 
briefly review various proposed physical realizations. 

\section{Perturbative results and strong coupling limit}
\begin{figure}
\begin{center}
\includegraphics[width=1.0\linewidth]{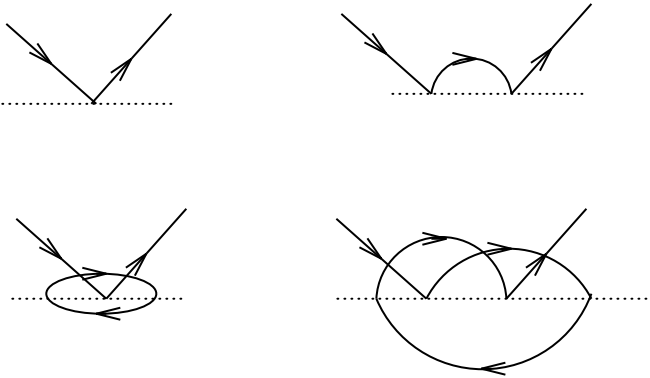}
\caption{Feynman diagrams contributing to the $\beta$-function 
to $O(\lambda^4)$.}
\label{fig:beta}
\end{center}
\end{figure}

It is relatively straightforward to calculate the $\beta$-function, 
which gives the change in effective coupling constant, $\lambda$, 
as the bandwith of the fermions is lowered, integrating out 
modes away from the Fermi surface. Although a pseudo-fermion 
representation is sometimes introduced for the impurity spin, 
it is simpler to calculate directly 
 expectation values of time-ordered products 
of spin operators in the non-interacting groundstate, using 
the spin commutation relations and $\vec S\cdot \vec S = S(S+1){\bf I}$, 
where $\bf I$ is the identity matrix. 
The spin operators have no time-dependence in this non-interacting 
groundstate but the time-ordering introduces some minus signs 
due to the non-commutation of the spin operators.  That is 
the time-ordered product:
\be{\cal T}[ S^a(t_1)S^b(t_2)]=\theta (t_1-t_2)S^aS^b
+\theta (t_2-t_1)S^bS^a.\ee
Note that the operators have no time-dependence on the right 
hand side of this equation.  In the S=1/2 case we may use:
\be S^aS^b=(1/4)\delta^{ab}{\bf I}+(i/2)\epsilon^{abc}S^c.\ee
Combining these minus signs with the free fermion propogators 
which can be conveniently written in position space at $x=0$ 
it is straightforward to calculate the $\beta$-function.  
The diagrams, to fourth order,
 are shown in Fig. (\ref{fig:beta}), 
and the result is:\cite{Abrikosov,Fowler}
\be
d\lambda /d(\ln D')=-[\lambda^2-(k/2)\lambda^3+ O(\lambda^4).]
\label{beta}\ee
[See App. B of Ref. (\onlinecite{Affleck3}) for a
quick derivation using the operator product expansion.]
The lowest order term, essentially found by Kondo, tells 
us that as we lower the cut-off, $D$, a small antiferromagnetic 
($\lambda >0$) 
 effective coupling begins to increase. Keeping only this 
leading term, we obtain the effective coupling:
\be \lambda (D') \approx {\lambda_0\over 1-\lambda_0 \ln (D/D')},\ee
where $\lambda_0$ is the bare coupling constant and $D'$ 
is the reduced bandwidth. This simply tells us that the 
effective coupling becomes large at $D'\approx T_K$, defined 
in Eq. (\ref{T_Kdef}), in the antiferromagnetic case, 
$\lambda >0$. 
Conversely, in the ferromagnetic 
case, $\lambda <0$, the effective coupling goes to 
zero at low energy scales.  
In the ferromagnetic case, a small bare 
coupling just keeps on getting smaller so that we never 
need to consider any terms beyond the first one in the 
$\beta$-function.  In this ferromagnetic case, the low 
energy fixed point corresponds to an impurity spin 
which decouples from the non-interacting conduction electrons. 
Conversely, in the antiferromagnetic 
case, at energy scales of 
$O(T_K)$ where $\lambda $ becomes O(1) it is not sufficient 
to keep only the leading term in the $\beta$-function. In fact, 
normally all high order terms become important at this 
stage and we lose control over the calculation. This is 
what makes the simple assertion that, in some sense, $\lambda \to \infty$ 
a non-obvious assumption in the simple, $k=1$ case. 

However, 
something very nice and very special happens 
in the antiferromagnetic case, when $k\gg 1$, as 
observed by Nozi\`eres and Blandin.\cite{Nozieres2}
 If we keep only the first 
two terms then the $\beta$ function vanishes at a fixed point coupling:
\be \lambda_c\approx 2/k.\label{lambda_c}\ee 
The third term in the $\beta$-function has no powers of $k$.  This 
can be seen from Fig. (\ref{fig:beta}), observing that the 
powers of $k$ come from fermion loops. Therefore, while the 
first two terms in the $\beta$-function are $O(1/k^2)$ at 
the fixed point, the third term is $O(1/k^4)$. In fact, all 
higher order terms are negligible compared to the first two 
at this fixed point at large $k$. 
Thus, as we reduce the bandwidth, the effective coupling flows 
to an asymptotic small, non-zero value at low energies. This 
correponds to a non-trivial critical behavior, different than 
the trivial strong coupling behavior ocurring for $k=1$. 
One way of seeing this is to observe that the slope of the 
$\beta$-function at the critical point is:
\be (d\beta /d\lambda )_{\lambda_c}\approx -2/k.\ee
At this intermediate coupling fixed point, we expect to be 
able to introduce a new basis of local operators and corresponding 
coupling constants. i.e.
\be H_{eff} = H_c + (\lambda -\lambda_c){\cal O} + \ldots \ee
Here $H_c$ is the fixed point Hamiltonian (i.e. free fermions 
together with a conformally invariant BC) and ${\cal O}$ is 
the leading irrelevant operator at the fixed point. The fact 
that the corresponding coupling constant, $\lambda -\lambda_c$ has 
the $\beta$-function:
\be d(\lambda -\lambda_c)/d\ln D = ((2/k)(\lambda -\lambda_c) 
+O[(\lambda -\lambda_c)^2],\ee
tells us that the corresponding operator, ${\cal O}$ has 
a non-trivial scaling dimension of $1+2/k$ and therefore 
this is not a Fermi liquid fixed point.  (At Fermi liquid 
fixed points all operators can be written in terms of free 
fermion fields and therefore have scaling dimensions 
which are integers or 1/2-integers.  Furthermore 
$H_{eff}$ only contains operators with even numbers of 
fermions which therefore have integer dimensions.) 

Further insight into the situation can be obtained by considering 
the strong coupling limit. Let us first consider the case $S=1/2$, $k=1$.
It is convenient to consider a tight-binding model with 
a hopping term $t$ in the limit $t<<J$ with the Kondo 
interaction on the site $0$ only. Thus we consider a Hamiltonian:
\be H = -t\sum_{<i,j>}\psi^\dagger_i\psi_j + 
J\psi^\dagger_0{\vec \sigma \over 2}\psi_0 \cdot \vec S.\ee
The first sum is over nearest neighbors on some lattice.  (The 
details of the lattice are unimportant.) We consider 
the case of large {\it antiferromagnetic} Kondo coupling, $J>>t>0$ .
The groundstate must have exactly one electron at the origin 
which forms a singlet with the impurity spin. Adding or 
removing one electron from the origin costs a large energy of $O(J)$. 
All the other electrons can do anything they like as long as 
they stay away from the origin. Thus, for a small non-zero $t$, 
they will form a free electron groundstate on all sites but the origin.
Low energy excitations simply correspond to the usual excitations 
of such a free fermion system with a vanishing boundary condition 
at the origin. In a spherically symmetric version of the problem, 
this BC corresponds to a $\pi /2$ phase shift in the s-wave channel. 

It is important to consider the self-consistency of this strong coupling 
fixed point.  Is it really plausible that $\lambda \to \infty$ 
under renormalization? i.e. is the fixed point stable?  We can 
answer this last question by considering all possible operators 
which could appear in the effective Hamiltonian at the fixed point 
which might potentially destabilize it. Importantly, these 
operators do not involve the impurity spin since it is screened 
and does not appear in the low energy $H_{eff}$. We must 
form operators purely from the electron operators and these 
can be shown to be irrelevant at this fixed point. [Actually, 
they are only strictly irrelevant when exact particle-hole 
symmetry is present.  Otherwise there is one exactly marginal 
operator.  However, it is not important for our purposes.]

\begin{figure}
\begin{center}
\includegraphics[width=1.0\linewidth]{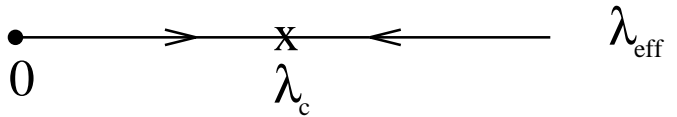}
\caption{ Renormalization group flow for the overscreened 
Kondo models.}
\label{fig:flow}
\end{center}
\end{figure}

Now consider the case of general $S$ and $k$ and again 
assume $J>>t$. Now the groundstate will have $k$ electrons 
at the origin, one from each channel. They will lock into 
a spin $k/2$ configuration. This spin $k/2$ couples 
antiferromagnetically to the impurity of spin $S$. The 
resulting groundstate will have a residual impurity spin 
of magnitude $|S-k/2|$.  This 
is non-zero in general, except in the case $S=k/2$. This 
is referred to as underscreening when $k/2<S$ and 
overscreening when $k/2>S$.  A crucial question is whether 
this effective spin couples 
ferromagnetically or anti-ferromagnetically to the 
conduction electrons. This is a straightforward calculation 
and the sign can actually be deduced from the well-known 
result that the induced Heisenberg exchange from the Hubbard 
model is antiferromagnetic. For similar reasons, 
the induced exchange interaction between the spin of the electrons 
 trapped at $0$ and the electron spins on the surrounding 
sites is antiferromagnetic.  However, this is not the end of the story.
We must consider the sign of the projection of the 
electron spin at the origin onto the effective 
spin (of size $|S-k/2|$).  This is positive in the 
overscreened case, $k/2>S$ and negative in the underscreened case, 
$S>k/2$. Thus the effective Kondo interaction between 
the over (or under) screened spin at the origin and 
the conduction electron spins on the surrounding sites 
is antiferromagnetic in the overscreened case (but ferromagnetic 
in the underscreened case).  This effective Kondo coupling is of order:
\be J_{eff} \propto t^2/J,\ee
which is $\ll 1$ in magnitude, for strong coupling. Thus 
we may analyse its RG flow using weak coupling perturbation theory, 
Eq. (\ref{beta}). 
The conclusion is that the strong coupling 
fixed point is  stable in the underscreened case, where the
partially screened spin of size $S-k/2$ asymptotically 
decouples from the free conduction electrons.  On the other 
hand, the strong coupling fixed point is unstable in the overscreened 
case where the antiferromagnetic Kondo coupling of the overscreened 
effective spin to the surrounding conduction electrons is 
relevant. We can summarize this situation by the RG flow 
diagram in Fig. (\ref{fig:flow}). Both zero coupling and strong 
coupling fixed points are unstable and the RG flows 
from either weak or strong coupling go to the intermediate 
coupling fixed point at low energies. These arguments don't 
depend on $k$ being large and we expect them to be 
qualitatively correct in general. Thus we conclude 
that the multi-channel Kondo model has a non-Fermi 
liquid stable low energy fixed point whenever $k>2S$. 

Unfortunately, these arguments don't tell us much about 
the low energy physics, except in the case $k>>1$ where 
some information can be extracted from weak coupling 
perturbation theory since $\lambda_c<<1$ in that case. 
Therefore, we turn to more powerful methods. 
 
\section{1 dimensional physics}
The Kondo model is fundamentally 1-dimensional.  This is a fortunate
observation because there are special techniques, applicable 
only in 1 dimension (1D), which can be brought to bear on the problem. 
To derive the 1-dimensionality in a simple way, we begin 
with a 3 dimensional model with a spherically symmetric dispersioin 
relation and assume that 
the Kondo interaction is a spatial $\delta$-function. If we then 
expand the electron operators in spherical harmonics, only the 
s-wave harmonic couples to the impurity.  The s-wave theory is 
equivalent to a 1D model defined on the 1/2-line $x>0$. Upon 
linearizing the dispersion relation, valid for weak Kondo coupling 
where we are only concerned with low energy states, the Hamiltonian
becomes:
\bea H &=& {iv_F\over 2\pi}\int_0^\infty  dx\left[\psi_L^\dagger {d\over dx}\psi_L
-\psi_R^\dagger {d\over dx}\psi_R\right]\nonumber \\
 &&+v_F\lambda \psi_L^\dagger (0){\vec \sigma \over 2}
\psi_L(0)\cdot \vec S,\eea
with the BC:
\be \psi_R(0)=\psi_L(0).\ee
[See, for example, App. A of Rev. (\onlinecite{Affleck3}) 
for a detailed derivation.]
This corresponds to a relativistic Dirac fermion, on the 1/2-line, interacting 
with the impurity spin at the origin. (We henceforth 
generally set the Fermi velocity, 
$v_F$, which plays the role of velocity of light, 
to 1.) 

A phase shift by $\delta$ (at all 
wave-vectors) corresponds to the BC:
\be \psi_R(0)= e^{2i\delta }\psi_L(0).\ee
In particular, for $\delta =\pi /2$ we get a new effective BC at 
the strong coupling fixed point:
\be \psi_R(0)=-\psi_L(0).\label{nbc}\ee
For any finite coupling, the phase shift would vary with $k$ but 
in the strong coupling limit, in a theory with a reduced bandwidth 
from RG transformations, we may consider the phase shift to be $\pi /2$ 
at all $k$. We then obtain this simple change of BC's.  So, the 
low energy effective Hamiltonian in the Fermi liquid case is simply:
\be H = {i\over 2\pi}\int_0^\infty  dx\left[\psi_L^\dagger {d\over dx}\psi_L
-\psi_R^\dagger {d\over dx}\psi_R\right],\ee
with the new BC of Eq. (\ref{nbc}). The impurity spin has disappeared 
from the low energy Hamiltonian and we are left with free fermions 
with a modified BC as the fixed point Hamiltonian. 

Starting with a
weak bare coupling, the only leading irrelevant operator can be 
shown to be:\cite{Affleck1}
\be H_{int} = {1\over T_K}\vec J (0)^2,\label{irr}\ee
where, 
\be \vec J(0)\equiv \psi^\dagger (0){\vec \sigma \over 2}\psi (0),\ee
and $T_K$ appears here simply as a coupling constant. The free fermion propogator is:
\be <\psi (\tau ,0)\psi^\dagger (0,0)>={1\over \tau}.\ee
Consequently the fermion operator has an RG scaling dimension of 1/2 
and the $\vec J(0)^2$ interaction has dimension 2.  The corresponding 
coupling constant thus has dimensions of inverse energy which 
justifies our calling it $1/T_K$ in Eq. (\ref{irr}).  [It follows 
from standard scaling arguments that this coupling constant, 
with dimensions of inverse energy should be of O($T_K$).  That 
it should be precisely $1/T_K$ is just a matter of definition 
of $T_K$.]  It is straightforward to do perturbation theory 
in $1/T_K$ when calculating low energy (eg. low temperature) quantities. 
This generates a series in $T/T_K$ (or more generally $E/T_K$ where $E$ 
is the energy at which a physical quantity is being calculated.)

Actually, there is another operator allowed in $H_{eff}$ when 
particle-hole symmetry is broken, 
$\psi^{i \alpha \dagger}(0)\psi_{i \alpha}(0)$. This 
has dimension 1 and is marginal.  However it can be 
shown to be strictly marginal; i.e. it does not renormalize at all. 
It generally has a small value and can be ignored.\cite{Affleck3}

In this simple case the new BC at the low energy fixed point is 
the trivial one of Eq. (\ref{nbc}). 
This turns out to be 
a fortuitously simple example of a much more general phenomenon
to which we now turn. 

\section{conformally invariant boundary conditions}
In quantum field theory in 1 space and 1 time dimensions, it is 
sometimes convenient to combine space and imaginary time ($\tau$) variables 
into a complex co-ordinate:
\be z\equiv \tau + ix.\ee
The conformal group is the set of analytic tranformations:
\be z\to w(z).\ee
The fact that this group is infinite dimensional in 2 space-time 
dimensions leads to many remarkable results. 
A quantum field theory defined on the half line, $x>0$, i.e. 
the upper half $z$-plane in the space-time picture, cannot 
be invariant under the full conformal group.  However, it 
{\it can} be invariant under the infinite sub-group which 
leaves the real axis (i.e. the boundary) invariant:
\be w(\tau )^*=w(\tau ).\ee
These transformations include time-translations, $z\to z+\tau_0$ scale 
transformations, $z\to az$ ($a\in R^+$) and an infinite set of 
others. Cardy developed powerful methods for classifying, discovering 
and studying all conformally invariant BC's.\cite{Cardy1}

An essential point to grasp about Cardy's general theory of conformally 
invariant BC's is that generally it is not possible to 
state explicitly what the BC is.  Fortunately this is not neccessary and 
all physical implications of a given BC can be extracted by 
ingenious indirect methods. It is useful to consider a complete set 
of conformally invariant BC's for a specified 
bulk conformal field theory (CFT).  For the Kondo model the 
bulk CFT is simply the free fermion Hamiltonian (with 2 spin 
components and $k$ channels.)  

The study of the finite 
size spectrum (FSS) plays a central role in conformal field theory. 
In the case of boundary CFT (BCFT) it is very useful to study 
the FSS on a strip of length $l$ with 
conformally invariant BC's $A$ and $B$ at $x=0$ and $x=l$. 
The complete set of finite size spectra (for 
a given bulk CFT) with an arbitrary pair of conformally 
invariant BC's actually encodes {\it all} information about all the 
boundary conditions. For instance, all Green's functions 
on the semi-infinite line with an arbitrary BC at $x=0$
can then be calculated both close to and far from the boundary 
for an arbitrary BC.  (Here we assume that the bulk 
Green's functions, in the absence of a boundary, are already know.) 

The method that Ludwig and I used\cite{Affleck2}
 to find the correct BC's 
corresponding to the overscreened Kondo models was actually 
based on first finding the FSS by a 
plausible conjecture. Apart from its 
fundamental role in BCFT, this FSS is also 
important because it is commonly calculated by Wilson's 
numerical renormalization group (NRG) method\cite{Wilson} 
for Kondo models.  Comparisons
 of the numerical  FSS with the BCFT prediction provides 
a powerful check on a conjectured BC.\cite{Affleck7}

Let us first consider the FSS with the BC's:
\bea \psi_L(0)&=&\psi_R(0)\nonumber \\
\psi_L(l)&=&-\psi_R(l).\label{bctriv}\eea
Since these BC's are true at all times, and since 
$\psi_L$ is a function only of $t+x$ while 
$\psi_R$ is a function only of $t-x$, it 
is possible to regard the right movers as the reflection 
of the left movers about either boundary. i.e.
\be \psi_R(x)=\psi_L(-x)=-\psi_L(2l-x)\ \  (0<x<l).\ee
Thus we may equivalently work with left-moving fermions 
only on a circle of radius $2l$ with anti-periodic BC's.
The corresponding free fermion FSS can be decomposed into 
multiplets of spin, $SU(2)$ and channel symmetry, $SU(k)$. 
Furthermore all states have a definite charge, as measured 
from that of the ground state. In fact, using 
conformal field theory techniques it is possible 
to decompose the spectrum into charge, spin and 
channel factors in a much more powerful way.\cite{Affleck2} Any 
state can be regarded as a produce of spin, channel 
and charge states with the excitation energy 
a sum of spin, channel and charge energies. Furthermore, 
all spin states can be grouped into a finite number of 
``conformal towers''.  These have the properties that 
the energies of all states in a conformal tower 
have the form:
\be E-E_0={\pi \over l}\left[\Delta + n\right],\ee
where $n$ is a non-negative integer 
which generally takes all values from $0$ to
$\infty$ in each conformal tower and $\Delta$ 
is a non-negative real number which is fixed 
for a given conformal tower. The spin quantum 
numbers of the infinite set of states in 
a given conformal tower are not fixed but range 
from a lowest value, $j$ to infinity. The 
spin conformal towers are conveniently labelled 
by the spin of the lowest energy state in the tower. 
[A similar structure also exists for channel 
and charge conformal towers but is not relevant to the 
present discussion.]  It turns out that, for $k$ channels 
of fermions, there is a total of $k+1$ spin conformal 
towers labelled by spins $j=0, 1/2, \ldots k/2$. 

 We now consider the FSS 
that occurs when we maintain the trivial BC of 
Eq. (\ref{bctriv}) at $x=l$ but introduce a non-trivial 
BC at $x=0$ corresponding to the Kondo model fixed point for 
spin $S$ and $k$ channels.  We found that the 
FSS in this case can be obtained from the spectrum 
with the free BC's of Eq. (\ref{bctriv}) by 
application of a general method introduced by 
Cardy (in this boundary context) and known as ``fusion''. 
The fusion rules give an algorithm for generating 
a new spectrum.  In this spin context the fusions rules  are 
a generalization of the ordinary angular 
momentum addition rules. The algorithm states 
that for each spin conformal tower occuring 
in the spectrum with the BC's of Eq. (
\ref{bctriv}) with spin $j$, a set of 
conformal towers occur in the Kondo spectrum 
with spin 
\be j'=|j-S|, |j-S|+1, \ldots ,\hbox{min}(j+S,k-j-S).\label{fusion}\ee
Here the last quantity represents the minimum of 
$j+S$ and $k-j-S$. To take a simple example, 
for $S=1/2$, $k=2$, the 3 conformal towers have $j=0$, $1/2$, and $1$.
The fusion rules replace these conformal towers by:
\bea 0&\to& 1/2\nonumber \\
1/2&\to& 0,1\nonumber \\
1 &\to & 1/2.\eea
In general, in the spectrum for the trivial BC's 
of Eq. (\ref{bctriv}) each time the spin $j$ conformal 
tower from the spin sector occurs together with some 
product of channel and charge conformal towers 
we must replace it by a sum of new products of conformal 
towers with $j$ replaced by the set of values of $j'$ 
given in Eq. (\ref{fusion}).
This finite size spectrum was compared to NRG results, 
in the case $k=2$, $S=1/2$ with excellent results.\cite{Affleck7}

Generalizing this approach, we can generate spectra corresponding 
to any pair from a complete set of BC's corresponding 
to a spin $S$ impurity at $x=0$ and a spin $S'$ impurity 
at $x=l$ with $S$, $S'\leq k/2$. Furthermore, this 
represents a complete set of spectra which contains 
enough information to determine the boundary Green's 
functions on the semi-infinite line with a Kondo 
BC at $x=0$. 

Note that we have turned the intuitive physical idea 
that the spin-$S$ impurity is somehow screened by 
the conduction electrons into a precise algorithm 
which gives the FSS and all other critical properties. 

Also note that since the spin conformal towers 
are labelled by spins from $j=0$ to $j=k/2$ only, 
in the undersceened case, $S<k/2$, it is not 
possible to construct the new spectrum as outlined above. 
In the underscreened  case $S$ is replaced by $k/2$ in 
Eq. (\ref{fusion}) and a decoupled impurity of size 
$S_{imp} = S-k/2$ remains in the low energy spectrum. 
It can be shown that fusion with the maximal spin conformal 
tower, $j=k/2$, always gives a trivial spectrum corresonding 
to the new BC:
\be \psi_R(0)=-\psi_L(0).\ee
Thus we get a Fermi liquid groundstate with a decoupled 
impurity spin, consistent with the naive strong coupling 
picture discussed in the previous section. 

Without attempting to give more details on the method, 
we now summarize some of the results which emerge from 
this BCFT approach. 

One of the most intriquing results is an exact formula 
for the equal-time correlation function of 
the left and right moving fermions:\cite{Ludwig1}
\be \langle \psi^{\dagger i\alpha}_L(x)\psi_{Rj\beta}(x)\rangle =
{\cos [\pi (2S+1)/(2+k)]\over \cos [\pi /(2+k)]}\cdot {\delta^i_j\delta^\alpha_\beta \over 2x}.
\label{GF}\ee
(Here, and in the formulas below it is assumed that $S\leq k/2$.  Otherwise 
$S$ must be replaced by $k/2$.) The free bulk Green's function is:
\be \langle \psi^{\dagger i\alpha}_L(x)\psi_{Lj\beta}(x)\rangle =
{\delta^i_j\delta^\alpha_\beta \over 2x}.
\label{GFf}\ee
In the exactly screened, or overscreened case, $S=k/2$, simple algebra 
shows that Eq. (\ref{GF}) reduces to:
\be \langle \psi^{\dagger i\alpha}_L(x)\psi_{Rj\beta}(x)\rangle =-
{\delta^i_j\delta^\alpha_\beta \over 2x}.
\label{GFm}\ee
Comparing to Eq. (\ref{GFf}), we see that the change of sign 
just reflects the Fermi liquid BC, $\psi_R(0)=-\psi_L(0)$, i.e. 
the $\pi /2$ phase shift. In general, we may write:
\be \langle \psi^{\dagger i\alpha}_L(x)\psi_{Rj\beta}(x)\rangle =S^{(1)}\cdot 
{\delta^i_j\delta^\alpha_\beta \over 2x},
\label{GFG}\ee
where $S^{(1)}$ is the $S$-matrix amplitude for an incoming electron 
to scatter into 1 electron (and no holes) at zero temperature 
and right at the Fermi surface. In the Fermi liquid 
case $|S^{(1)}|=1$. In the non-Fermi liquid case, $|S^{(1)}|<1$.
 Since the $S$-matrix must be unitary when all processes are 
included, this implies that the probability of producing 
multi-particle final states is finite, even at $T=0$ and right 
at the Fermi surface in the overscreened case. This is 
perhaps the clearest demonstration that the overscreened fixed 
points correspond to non-Fermi liquid groundstates. From 
this single fermion Green's function, in the presence of a single 
impurity at the origin, we can find the self-energy in the case 
of a dilute random array of impurities, and hence the resistivity.  
This gives the $T=0$ dc resistivity:
\be \rho (0) = {3n_i\over k\pi (e\nu v_F)^2}\left[{1-S^{(1)}\over 2}\right] .\ee
Here $n_i$ is the density of impurities. 
In the case $S^{(1)}=-1$, ($\pi /2$ phase shift) we obtain the ``unitary limit''
resistivity.  In non-Fermi liquid cases the resistivity is reduced. A useful 
check is to consider the large $k$ limit. In this case, $S^{(1)}\to 1$ and 
we recover the Green's function with the original BC $\psi_R(0)=\psi_L(0)$. 
In this limit, the resistivity goes to zero.  This corresponds to a weak 
critical Kondo coupling, $\lambda_c=2/k\to 0$, from Eq. (\ref{lambda_c}).
The case $k=2$, $S=1/2$ (or, in general $k=4S$) is especially interesting 
because now $S^{(1)}=0$.  In this case all the scattering is multi-particle 
with zero amplitude for a single electron to scatter into a single electron!

It is also possible to calculate the leading $T$ dependence of the resistivity 
at low $T$. \cite{Affleck5} 
This requires doing perturbation theory in the leading irrelevant 
coupling constant at the low energy fixed point. Our solution for the 
non-Fermi liquid BC determines this coupling constant to have 
renormalization group eigenvalue:
\be y=2/(2+k).\ee
  It then follows by a standard scaling 
argument that this coupling constant is $1/T_K^y$ times a constant 
of order 1.  Thus the resistivity, at low $T$ goes as:
\be \rho (0) \to  {3n_i\over k\pi (e\nu v_F)^2}\left[{1-S^{(1)}\over 2}\right]
\left[1+c(T/T_K)^y\right] .\ee
Here $c$ is a constant of $O(1)$. In the Fermi liquid case, the leading 
irrelevant operator is $\vec J^2(0)$ as mentioned in the previous 
section and the corresponding coupling constant has dimension $1$.  
However, in this case it contributes to the resistivity only  
beginning in second order or perturbation theory giving:
\be \rho (0) \to  {3n_i\over k\pi (e\nu v_F)^2}
\left[1-c(T/T_K)^2 \right] .\ee
In particular the $T$ dependence of $\rho$ is quadratic for $k=1$ 
but square root for $k=2$ and $S=1/2$. 

The impurity entropy also shows interesting behavior.  In the Fermi liquid
case, at $T=0$, this is simply:
\be S_{imp}(0)=2S_{eff}+1,\ee
where $S_{eff}=S-k/2$, is the size of the underscreened spin.  (This becomes 
zero for exact screening.) This is simply the entropy of a decoupled 
spin of size $S_{eff}$ and reflects the asymptotic decoupling 
of the electronic degrees of freedom from the partially screened spin 
at $T\to 0$. On the other hand, from the 
non-Fermi liquid BC for overscreening, we find:\cite{Affleck4}
\be
S_{imp}(0) = {\sin [\pi (2S+1)/(2+k)]\over \sin [\pi /(2+k)]}.\ee
This is always between $1$ and $2S+1$, corresponding to non-trivial 
partial screening of the impurity spin. It is 
generally {\it not} an integer. This result was 
first obtained from the Bethe ansatz solution of the multi-channel 
Kondo problem\cite{Andrei,Wiegmann}
 and thus provides a valuable check on the BCFT approach. 

The $T$-dependence of the impurity entropy at low $T$, i.e. the specific heat, 
can also be calculated from lowest order perturbation theory 
in the irrelevant operator.  It can be found at all $T$ from the 
Bethe ansatz, giving compatible results. The same is true of 
the impurity susceptibility.  These quantities exhibit fractional 
power law behavior in the non-fermi liquid case:\cite{Affleck3}
\bea C_{imp}(T)&\to& (T/T_K)^{2y}\nonumber \\
\chi_{imp}(T) &\to & (T/T_K)^{2y -1}.\eea
The behavior is different in the special case $k=2$, $S=1/2$ where 
$2y=1$.  We now obtain:
\bea C_{imp}(T)&\to& (T/T_K)\ln (T_K/T)\nonumber \\
\chi_{imp}(T) &\to &\ln (T_K/T) .\eea

\section{symmetry breaking perturbations and physical realizations}
The multi-channel Kondo model has an exact $SU(k)$ channel symmetry 
in addition to exact $SU(2)$ spin symmetry. In possible physical 
realizations this symmetry is usually partly broken by extra 
terms in the Hamiltonian. 

For instance, the channel symmetry might be broken with 1 channel 
coupling more strongly to the impurity that the rest. The interaction 
term of Eq. (\ref{hamk}) is then modified to:
\begin{equation}
H=\sum_{\vec{k}\alpha ,i}\psi^{\dagger\alpha i}_{\vec{k}}
\psi_{\vec{k}\alpha i}\epsilon(k)+\vec{S}\cdot\sum_{\vec{k}\vec{k'},i}J_i
\psi^{\dagger i}_{\vec k} \frac{\vec{\sigma}}{2}\psi_{\vec{k'} i}
\label{haman}\ee
It is easy to see how this effects the $\beta$-functions for the $k$
different couplings, $\lambda_i$. The quadratic term is a self-interaction 
whereas the cubic term contains a Fermion loop which implies a sum over 
all channels:
\be
d\lambda_i /d(\ln D')=-\lambda_i^2+(k/2)\lambda_i\sum_{j=1}^k\lambda_j^2+ O(\lambda^4).]
\label{betaan}\ee
If one of the coupling constants, say $\lambda_1$, is initially larger than the rest, 
then Eq. (\ref{betaan}) implies that it grows faster than the rest.  Indeed, the
other couplings eventually start to {\it decrease} under renormalization. 
This begins to happen when the $j=1$ term in the cubic term in $\beta$-function for $\lambda_i$ 
(with $i>1$) starts to dominate over the quadratic (and other) terms. The simple 
physical picture, in the $S=1/2$ case, is that one channel screens the impurity 
and the other $k-1$ channels asymptotically decouple. Thus we effectively get 
a single channel fixed point which is of Fermi liquid type.

We note in passing that this observation may be important in explaining 
why single-channel Kondo behavior is apparently observed quite commonly 
in metals with dilute magnetic impurities. Our discussion so far 
has always assumed that the Kondo interaction was a spatial $\delta$-function, 
so that only the s-wave degrees of freedom of the electrons couple to the impurity. 
Allowing for a finite range interaction, the other harmonics will couple as well. 
We can still map to a 1-dimensional model but with many channels, all 
with different Kondo couplings. Since the s-wave will generally couple 
most strongly the previous naive 1-channel analysis still applies. 

To actually verify that channel symmetry breaking is a relevant 
perturbation at the non-Fermi liquid fixed point it is 
neccessary to calculate the scaling dimension of 
the most relevant new operator which appears in the Hamiltonian 
when channel symmetry is broken.  The conclusion is\cite{Affleck5} 
that this operator has dimension, $\Delta =k/(2+k)$, and 
thus is relevant.

Another possible type of symmetry breaking interaction is 
exchange anisotropy which preserves channel symmetry but 
breaks ordinary spin-rotation symmetry. This is found 
to be irrelevant at both Fermi and non-Fermi liquid 
fixed points.\cite{Affleck5} 

Finally, we can consider a local magnetic field, acting 
only on the impurity spin.  This is found to be 
relevant at the non-Fermi liquid fixed points.\cite{Affleck5}  In 
particular, the corresponding boundary operator 
has dimension $2/(2+k)$. 

There have been several proposals for experimental 
realizations of non-Fermi liquid Kondo behavior, 
primarily in the simplest case $k=2$, $S=1/2$. 
An early proposal
involves an atom in a double well potential coupled 
to conduction electrons.\cite{2LS} In this case, the two 
different locations of the atom correspond to 
 $S^z=\pm 1/2$ for the impurity ``spin''. A 
simplified treatment keeps only 2 angular momentum 
states of the conduction electrons, the s-wave 
and the p-wave (with aximuthal component, $m=0$). 
These two angular momentum components of the 
conduction electrons play the role of conduction 
electron spin.  A very anisotropic Kondo interaction 
exists between the conduction electrons and the atom. 
The $\sigma^zS^z$ term represents electrons 
in either channel scattering off the atom in 
either location.  the $\sigma^xS^x$ term 
represents ``electron-assisted tunnelling'' in 
which the atom tunnels from one location to 
the other while the scattering electron changes 
angular momentum states. It is crucial to 
note that the real electron spin is assumed to be a completely 
passive quantum number in the scattering/tunnelling process. 
Thus it plays the role of the ``channel''
 in the 2-channel Kondo model. Since spin anisotropy 
is irrelevant, as mentioned above, this model would 
renormalize to the non-Fermi liquid fixed point. However, 
there are 2 other interactions which must be included. 
One of these is direct tunnelling of the atom between 
the 2 locations (without any electron involvement).  This 
corresonds to a term $\Delta \cdot S^x$ in the Hamiltonian, 
where $\Delta$ is a parameter with dimensions of energy.  
Finally, the energies of the atom at the 2 locations 
are generally unequal, leading to a term $\Delta_0\cdot S^z$. 
Both of these terms correspond to a local magnetic 
field acting on the impurity only, which is 
relevant, as mentioned above. Only if 
\be \Delta, \Delta_0\ll  T_K\label{cond}\ee
does  non-Fermi liquid behavior occur.  In that 
case we expect we expect to see approximate non-Fermi liquid 
behavior over an intermediate energy range:
\be \Delta ,\Delta_0<<E<<T_K.\ee
It is not obvious that the condition of Eq. (\ref{cond}) 
ever occurs in any real system although various proposals have been made. 
A relatively recent proposal involved electrons tunnelling 
through a nano-constriction.\cite{Ralph}
  The proposal is that, in some cases, 
the tunnelling goes through localized states in the nano-constriction 
which can exist in 2 different configurations, corresponding 
to ``impurity spin''. 
A closely related proposal involves crystal field states 
of an atom in a crystal.\cite{2LS}

Another pair of proposals involves the Coulomb blocade in 
a quantum dot. In one scenario\cite{Matveev} the voltage  on the 
quantum dot is tuned to a critical point where 2 
possible charges of the dot have the same energy. In 
this case the 2 charge states of the dot behave as 
the effective $S^z$ eigenstates. Again the electron 
spin plays the role of the 2 ``channels'' in 
the effective 2-channel Kondo model. An alternative 
scenario\cite{Oreg} involves a quantum dot where the voltage 
is such that there is an odd number of electrons 
on the dot which is then assumed to have a spin-1/2 
ground state.  This is closer to the original 
formulation of the Kondo model in which it is 
real electron spin which couples to the impurity.  
In the usual transport experiments, where 2 
normal leads are connected to 
the quantum dot, a single channel Kondo model 
is appropriate. This single channel correponds 
to the symmetric linear combination
formed from the 2 leads (``s-wave'').  In fact, 
even if a larger number of normal leads is connected 
to the dot, the single channel  Kondo model 
is still the appropriate description because 
only a single symmetrized channel, a symmetric linear 
combination formed from all the leads, couples 
to the  spin of the quantum dot.  A 2 channel 
model can be obtained if one of the leads is 
replaced by a mesoscopic sized island of electrons, 
a ``large dot''. 
In this case tunnelling of electrons onto 
or off of this island may be suppressed by its 
Coulomb blockade.  If only 
Kondo coupling terms are kept in the Hamiltonian 
which preserve the number of electrons 
on this island, it can acts as a second channel 
leading to 2 channel behavior. However, 
this requires that the finite-size level spacing 
of spin excitations of the island be small compared 
to the Kondo temperature which itself 
must be small compared to the Coulomb blockade energy. 
Furthermore, a gate voltage must be fine-tuned in 
order to make the Kondo couplings to the two leads 
equal. 

Thus, it remains an interesting open 
question which or not non-Fermi liquid behavior 
in a Kondo model has been or will soon be observed 
experimentally.

\begin{acknowledgements} 
I would like to thank my principal collaborator in this work, 
Andreas Ludwig.  This research was supported in part 
by the Canadian Institute for Advanced Research and NSERC of Canada.
\end{acknowledgements}

\end{document}